
%
%
%

\magnification = \magstep1


\input epsf
\input rotate

\font\sc=cmcsc10
\font\ninerm = cmr9
\font\ninebf = cmbx9
\font\nineit = cmti9
\font\ninei  = cmmi9
\font\ninesy = cmsy9
\font\ninett = cmtt9

\countdef\refcount=50 \refcount = 0
\countdef\eqcount=100 \eqcount = 0

\def\setrefnum#1{
   \advance\refcount by 1
   \count255 = 50 \advance\count255 by \refcount
   \countdef#1=\count255 #1=\refcount
   }
\def\seteqnum#1{
   \advance\eqcount by 1
   \count255 = 100 \advance\count255 by \eqcount
   \countdef#1=\count255 #1=\eqcount
   }

\newbox\BodyBox
\newbox\CaptionBox
\newbox\RotateBox

\def\makeanemptybody#1{
   \setbox\BodyBox = \vbox to #1 truecm {}
   }

\def\figure#1#2{
   \setbox\CaptionBox=\vtop{
      \textfont0=\eightrm \scriptfont0=\fiverm
      \textfont1=\eighti  \scriptfont1=\fivei
      \textfont2=\eightsy \scriptfont2=\fivesy
      \textfont3=\eightex
      \multiply \hsize by 2
      \divide \hsize by 3
      \noindent {\eightbf Figure~{#1}.} \eightrm #2
      }
   \topinsert
      \centerline{\box\BodyBox}
      \smallskip
      \centerline{\box\CaptionBox}
   \endinsert
   }


\def\revtit#1#2#3#4#5{``#1'' {\it #2} {\bf #3}, #4 (#5).}
\def\revbook#1#2#3{{\it #1}, #2 (#3).}
\def\author#1{{\sc #1},}

\def\mathinsert#1{\vtop{
   \divide\hsize by 8 \multiply\hsize by 5
   \divide\baselineskip by 6 \multiply\baselineskip by 5
   \noindent \it #1}}

\def\beginalgol{\vtop{
   \settabs 26 \columns \it \bgroup } }
\def\endalgol{\egroup}
\def\resword#1{{\bf #1}}


\font\titlefont = cmr10 scaled \magstep3
\def\titleline#1{\centerline{\titlefont #1}}

\def\begintitle{\setbox1=\vbox{%
   \null \vskip 0.5 truecm
   \baselineskip = 20pt
   \bgroup}}
\def\endtitle{\egroup \noindent\box1}

\def\name#1{\centerline{\ninerm #1}}
\def\address#1{\centerline{\nineit #1}}
\def\email#1{\centerline{\ninett #1}}

\def\beginauthors{\setbox1=\vbox{%
   \null \vskip 1 truecm
   \baselineskip = 10pt
   \bgroup}}
\def\endauthors{\vskip 1.5 truecm \null \egroup
   \noindent\box1}

\def\beginabstract{ \setbox1 = \vbox{%
   \advance\hsize by -4 truecm
   \baselineskip = 11 pt
   \ninerm
   \textfont0=\ninerm  \scriptfont0=\sevenrm
   \textfont1=\ninei   \scriptfont1=\seveni
   \textfont2=\ninesy  \scriptfont2=\sevensy
   \centerline{\ninebf Abstract}
   \medskip \bgroup \noindent }}
\def\endabstract{ \egroup \centerline{\box1} }

\def\today{$ \rm
   \ifcase\month\or January \or February \or March \or
   April \or May \or June \or July \or August \or
   September \or October \or November \or December \fi \;
   \number\day, \; \number\year $}

\def\preprint#1#2{\line{#1 \hfil INFN-ROM1 #2}}

\def\section#1{\vskip 1 truecm
   \noindent {\bf #1} \medskip}


\nopagenumbers

\preprint{July 22, 1994}{1040}
\line{\hfil cond-mat/9407092}

\begintitle
\titleline{Subminimal Paths on a Stochastic Graph}
\endtitle

\beginauthors
\name    {Giovanni Ferraro}
\address {Dipartimento di Fisica dell'Universit\`a di Roma ``La Sapienza''}
\address {and INFN -- Sezione di ROMA}
\email   {FERRAROG\%VAXRMA.hepnet@LBL.GOV}
\endauthors

\beginabstract
A simple model of a frustrated disordered system is presented.
Apart from the (very different) physical interpretation,
the model shares many features with that of
Sherrington-Kirkpatrick for spin glasses, but,
as a consequence of its relative
simplicity, its ground state can
be exactly determined by numerical methods. This fact allows us to test
experimentally some theoretical predictions, based on a specialization of
the ``cavity method'' developed for the SK model, which is
presently limited to a ``non-frustrated'' approximation, corresponding
to some extent to the replica-symmetric one for the SK model.
\endabstract

\vfill
\eject
\footline {\hfil \tenrm\folio \hfil}


\setrefnum\SK
\setrefnum\Parisi
\setrefnum\Overlap
\setrefnum\Ultra
\setrefnum\RFE
\setrefnum\Cavity
\setrefnum\Ruelle
\setrefnum\CPPS
\setrefnum\FH
\setrefnum\REM
\setrefnum\GREM
\setrefnum\Papa
\setrefnum\Barahona
\setrefnum\dyrpol
\setrefnum\PDG
\setrefnum\BS

\seteqnum\defpath
\seteqnum\defl
\seteqnum\defp
\seteqnum\defw
\seteqnum\greedy
\seteqnum\neighMST
\seteqnum\steepest
\seteqnum\nftype
\seteqnum\Ihc
\seteqnum\pexact
\seteqnum\approne
\seteqnum\lnresult
\seteqnum\phl
\seteqnum\lzero
\seteqnum\apprtwo
\seteqnum\FNrel
\seteqnum\FNinfty
\seteqnum\FNzero
\seteqnum\plzero
\seteqnum\lmean
\seteqnum\csrel
\seteqnum\fit


\line{}
\section{1.~Introduction}

\noindent
The Sherrington-Kirkpatrick model for spin glasses ([\number\SK],
SK model in the following) was proposed in 1975 as a simple and exactly
solvable model of a disordered frustrated system.  This is a
model with very long range interaction,
for which the mean field theory is supposed to hold exactly, and
therefore it should be easy to handle. But the hopes of the authors were
soon disappointed: the model is exactly solvable, but its solution
turned to be much more complex than what initially thought. Only in
1980 the work of Parisi [\number\Parisi]
and the complicated and mysterious ``breaking of the replica symmetry''
solved the problem, and only a few years later
the so called {\sl cavity method}
[\number\Overlap,\number\Ultra,\number\RFE,\number\Cavity]
revealed the physics hidden behind the intricate mechanism of the
replica formalism.

Two features appear to be essential in the model, both linked to a highly
non-trivial decomposition of the Boltzmann measure in pure states:

\item{1)} the pure states are organized in an ultrametric scheme
[\number\Overlap,\number\Ultra];

\item{2)} the free energies at each level of this organization are
Poissonian random variables with exponential density, and the levels of
organization are connected through a probability cascade {\sl \`a la} Ruelle
[\number\RFE,\number\Ruelle].

It is not yet clear whether this picture can be applied to other frustrated
disordered models, such as 3-dimensional spin glasses with nearest
neighbours interaction [\number\CPPS,\number\FH].\footnote{${}^{(*)}$}
{Models like the REM and the GREM of Derrida [\number\REM,\number\GREM],
though very instructive, are rather artificial.}
For this reason, we have considered a model which preserves many features of
the SK model, but is much simpler to be analyzed numerically.
Our hope is
twofold: to clarify if and how one can apply the same ``ultrametric''
picture as the SK model, and to study the nearest neighbours
interaction case, taking
advantage of a better numerical tractability. This letter
represents a first step in this direction.

\medskip

The outline of the paper is as follows. In section~2 we define the model,
in section~3 we clarify the concept of frustration in the context of our
model, in section~4 we propose two approximate analytical solutions of the
model, in section~5 we show some numerical results, as a test for the
analytical ones.

\section{2.~Definition of the model}

\noindent
We consider a set of points $V$, connected through links of a
given length, defining the distance between each pair of points of $V$.
We start by stating the problem with the greatest generality,
with the points of $V$ not necessarily
distributed in the real space, and their relative
distances not necessarily being the euclidean ones.
To be precise,
let us recall a few definitions. A {\sl weighted graph} is a triple
$ G=(V,A,d) $ where $V$ is a finite set of {\sl nodes} or {\sl vertices},
$ A \subset V \times V $ is the set of the {\sl arcs} ({\sl i.e.} the links)
which connect the
nodes of $V$ and the map $ d : A \to {\bf R}^+ $ gives the length
(or {\sl weight}) of an arc, {\sl i.e.} the distance between
the nodes of an (ordered) pair.
A {\sl path} of $n$ steps in the graph
starting from $u$ and leading to $v$ is a sequence of $n+1$ nodes
$ \{ u_i \}_{i=0}^n $ such that
$$
\eqalign{
u_0 = u \; , \quad u_n & = v \; ,
\quad u_i \not= u_j \; \; \forall \; i,j = 0 , \dots , n \; \; i\not=j \cr
& (u_i,u_{i+1}) \in A \; \; \forall \; i = 0 , \dots , n-1 \cr
} \eqno (\number\defpath)
$$
Let now $ C_{u,v} $ be the set of the paths joining $u$ to $v$,
irrespective of the number of their steps. We suppose that the graph is
connected, {\sl i.e.} that there is at least one path joining $u$ to $v$.
The length of paths in $ C_{u,v} $ is represented by a map
$ l $ defined on $ C_{u,v} $ in a natural way, {\sl i.e.}
$ l : C_{u,v} \to {\bf R}^+ $ such that
$$
c \in C_{u,v} \; , \; c = \{ u_i \}_{i=0}^n \; ,
\quad l[c] \equiv \sum_{i=0}^{n-1} d(u_i,u_{i+1})
\eqno (\number\defl)
$$
Our system has phase space
$$
\Omega = C_{s,t}
$$
with $s$ and $t$ fixed in advance, and Hamiltonian
$$
{\cal H} = l
$$
The ground state of the system is the shortest
path in $G$ between $s$ and $t$. Using
a standard algorithm, the Dijkstra's algorithm
(see for instance [\number\Papa] and references therein), it can be
determined in a time which grows not too fast with the size of the
problem, for any realization of the distance matrix $ d(u,v) $.
On the contrary, to find the ground state of the SK model is a so
called $NP$-complete problem (see [\number\Papa] and also
[\number\Barahona]) and therefore it is very unlike
that a ``good'' algorithm can be found for it.

The distances $ d(u,v) $ play the same
role of the couplings $J_{ij}$ among
the spins in a generic model of interacting Ising spins.
In the case of our interest
they will be random variables and will represent
the ``disorder'' of the model. How frustration arises in this context will
be discussed in the next section.
As anticipated, our main interest in studying the problem is to
gain informations about the behaviour of disordered and frustrated systems,
in a context which has a rather rich structure, but nevertheless is
simpler to handle than spin systems. Apart from this rather abstract
motivation, the study of our problem could be useful in the framework of
high-temperature expansion for disordered systems of interacting Ising
spins. Moreover, also the problem of directed polymers in a random medium
(see, for some recent results, the work in [\number\dyrpol] and references
therein) could be stated as a particular case of our problem. In fact, a
directed polymer is a path in a particular graph which has a
privileged direction of motion: that is to say, with the
previously introduced notation, that if the
link $ (u,v) \in A $ points in the privileged direction then
$ (v,u) \not\in A $ (the reversed link is absent). With our notation
the random medium is represented by a distance matrix written in terms of a
single random function $ \phi $, the local disorder,
$$
d (u,v ) = { \phi(u) + \phi(v) \over 2 }
$$
so that equation~(\number\defl) for the length of a path $c$ of
$n$ steps becomes
$$
l[c] = \sum_{i=0}^n \phi(u_i) - { \phi(u_0) + \phi(u_n) \over 2 }
$$
which is the usual potential energy term in the hamiltonian of a polymer in
a local disorder field, apart from a contribute vanishing in the limit
$ n \gg 1 $.

The model we shall consider in the following is defined by:

\item{1)} $ (u,v) \in A $ for every $ u,v \in V $, with $ u\not=v $;

\item{2)} the $ d(u,v) $'s are independent identically
distributed random variables (with $ d(u,v) = d(v,u) $).

\noindent
Statements~1) and~2) are the counterparts
of the defining characterization of
the SK model, the ones by which mean field theory turns out to hold
exactly. This analogy makes us confident that one could find an exact
solution for our model, although the meaning of
``mean field theory'' in our context is not clear.
Moreover, the fact that from a numerical point of view the problem
is much simpler to deal with than the SK model, lead us to hope
that the same happens also in an analytical approach.
We did not succeed in finding such an
exact solution, nevertheless thanks to the~1) and~2) above we can obtain an
approximate solution (see section~4) which fits well enough the numerical
data (see section~5).

We choose as probability distribution for the $d$
$$
p_\alpha (x) = ( \alpha+1 ) \cdot x^\alpha \quad , \quad x \in [ 0,1 ]
\eqno (\number\defp)
$$
In the analytical computation of section~4, $\alpha$ remains a free
parameter, whereas in the numerical simulation of section~5 we fix, just
for concreteness, $\alpha=0$ and $\alpha=1$, {\sl i.e.} we consider,
respectively,
uniform and linear probability distribution of the distances.

\section{3.~Frustration}

\noindent
It is not immediate to understand how the usual concept of frustration,
developed in the framework of spin systems, applies to this context. What
naturally characterizes frustration, in a context-free way, is the
existence of a lot of local minima, hampering the selection of the global
one. Although in our model we can not precisely identify the local minima,
we propose a generic definition of frustration, which easily
implies that our model is frustrated.

It is now useful to introduce another model, structurally
similar to ours.
Let us recall that a {\sl tree} is a connected graph without cycles
and a {\sl spanning tree} of a graph is a tree which has the same set of
nodes as the graph. The phase space of this second model is the set
${\cal S}T$ of all the spanning trees of $G$, and the Hamiltonian is the map
{\sl weight} $ w : {\cal S}T \to {\bf R}^+ $
naturally defined by
$$
T \in {\cal S}T \; , \; T = (V,A_T) \; ,
\quad w[T] \mkern 5mu \equiv \mkern -15mu \sum_{(u,v) \in A_T} d(u,v)
\eqno (\number\defw)
$$
In the following we shall call {\sl minimal spanning tree} ($MST$) the ground
state of this system. As a matter of fact, the Dijkstra's algorithm,
which determines the shortest path from $s$ to $t$,
gives also the shortest path from $s$ to any other $ u \in V $.
These paths form a tree, the so called {\sl shortest path tree} ($SPT$).
Thus, to a certain extent, the $SPT$ problem and the $MST$ problem are similar,
as their solutions are both represented by trees on the graph (they do not
coincide, however, as it is easily shown by a counterexample).
On the other hand the algorithms which solve them are rather different.
The Dijkstra's algorithm is somewhat tricky.
On the contrary, the algorithm which solves the $MST$ problem is much simpler:
$$
\beginalgol
\+ $ MST := \emptyset $ \resword{;} \quad $ A' := A $ \resword{;} \cr
\+ \resword{repeat} \cr
\+ & \resword{let} $a$ the arc in $A'$ of minimal length \resword{;} \cr
\+ & $ A' := A' \setminus \{ a \} $ \resword{;} \cr
\+ & \resword{if} $ MST+a $ is a tree
  \resword{then} $ MST := MST + a $ \resword{;} \cr
\+ \resword{until} $ A' = \emptyset $ \resword{;} \cr
\endalgol
\eqno (\number\greedy)
$$
This algorithm is called {\sl greedy} for obvious reasons.

Systems for which the greedy algorithm to find the ground state
works are called {\sl matroids}. Though this seems to be an informal
characterization, it turns to be equivalent to a purely algebraic one (see
[\number\Papa]).
For matroids there exists a natural way to give a notion of
neighbourhood between the points in the phase space, that for the spanning
trees of a graph results in:
$$
\mathinsert{If $ T \in {\cal S}T $ then the neighbours of $T$ are all the
$ T' \in {\cal S}T $ obtained from $T$ as follows: add an arc
to the tree $T$, producing a cycle; then delete one other arc on the cycle.}
\eqno (\number\neighMST)
$$
If we consider the local minima of the Hamiltonian $w$ with respect to this
``topology'', it is an amusing exercise
[\number\Papa, exercise~3 chap.~1]
to show that, apart from accidental degenerations, there exists a
unique local minimum, which therefore is a global one. Thus the following
{\sl steepest descent} algorithm is equivalent to the greedy one, in order
to determine the $MST$:
$$
\beginalgol
\+ \resword{let} $ T \in {\cal S}T $ \resword{;} \cr
\+ \resword{repeat} \cr
\+ & \resword{let} $T'$ the neighbour of $T$ of minimal weight \resword{;} \cr
\+ & \resword{if} $T'\not=T$ \resword{then} $ T := T' $ \resword{;} \cr
\+ \resword{until} $ T' = T $ \resword{;} \cr
\+ $ MST := T $ \resword{;} \cr
\endalgol
\eqno (\number\steepest)
$$
Now, it seems natural to consider the matroids as the archetypical example
of a non-frustrated system, in contrast with the frustrated ones, where, as
it is well known, the huge number of local minima causes the steepest
descent method to fail. We remark that, to adapt the
greedy algorithm to a spin system, it is necessary to orient subsequently
the spins of the system, so that the couplings with the already
oriented ones are satisfied.
It is not difficult to make the above procedure formal, and
to verify that it works if and only if for every $n$-tuple of lattice sites
$ \{ i_1, \dots, i_n \} $ the following condition holds:
$$
J_{i_1 i_2} \cdot J_{i_2 i_3} \cdots J_{i_n i_1} \ge 0 \; ,
$$
where $ J_{ij} $ is the coupling constant between the spin $\sigma_i$ and
the spin $\sigma_j$. This is nothing but the usual definition of absence of
frustration. In conclusion, two facts lead us to conjecture that matroids
are the most general non-frustrated systems:

\item{-} there exists a
unique local minimum, which therefore is a global one;

\item{-} the spin systems which are matroids are the non-frustrated ones,
in the usual meaning.

\noindent
{}From this point of view our model is frustrated, and the $MST$ problem
becomes
its non-frustrated counterpart.
To prove that our system is frustrated it is sufficient to show
that it is not a matroid, and this can be done in two ways:
either by its algebraic characterization,
or by showing with a counterexample that to be greedy
does not work. Roughly speaking, in order
to find the shortest path between $s$ and
$t$, it is not sufficient in general to proceed na\"\i vely, {\sl i.e.} by
choosing always the shortest arc when going out from a node.

\medskip

In the following, we shall fix our attention on the shortest path between
two fixed nodes in a graph, which we denote with $s$ and $t$. The size
of the graph, {\sl i.e.} the number of its nodes, will be denoted by $N$, and
we
shall be interested in the asymptotic properties in the limit
$ N \to \infty $. Moreover, as it will be clear in the
following, we shall study, together with the shortest path, also the paths
in the graph between $s$ and $t$ which are next to the shortest one,
{\sl i.e.} we order the paths by increasing length, and we shall be interested
not only in the first one, but also in the second one, the third one,
and so on.
These will be called ``subminimal'' or ``suboptimal'' paths, whereas the
shortest path is the minimal, {\sl i.e.} the optimal, one. We do not attempt to
give a rigorous definition of this notion, {\sl i.e.} we do not
say how close a path should be to the shortest one, to be called a
subminimal one. However, if we order by increasing length all the paths
between $s$ and $t$ in the graph, it is clear that the $h$-th one, with $h$
fixed, is always subminimal in the limit $N\to \infty$, since it becomes
{\sl infinitely} close to the shortest one. In fact,
as far as we could say about our problem, by a partial analytical solution
and a numerical one, all the shortest $k$ paths in the graph, with $k$
fixed, in the limit $ N \gg 1 $ share the same leading behaviour
{\sl versus} $N$ (see (\number\lnresult) below, and section~5). Thus we
could take as an informal characterization of the subminimal paths the
fact that they all have the same asymptotic behaviour {\sl versus} $N$, when
$N$ goes to infinity. It is clear that the number of subminimal paths
must grow with $N$, but we do not know at all how.

We believe that this characterization, though informal, is not trivial.
Roughly speaking, we do not know exactly which paths are the subminimal
ones, but we can argue which of them are surely not. In fact,
let us suppose we are
looking for the shortest path between $s$ and $t$ which visits also
{\sl any} other node in the graph. Such a path is the so called
{\sl Hamilton path}; to find it is a problem very close to the well
known {\sl travelling salesman problem} ($TSP$), which is a so called
{\sl $NP$-complete} problem (see [\number\Papa]), {\sl i.e.} so hard to solve
as to
find the ground state of the $SK$ model. On the other hand, our problem,
{\sl i.e.} the problem of finding the shortest path between two nodes in a
graph, is
one of the so called {\sl class-$P$} problems, and it is much easier to
solve numerically. It is commonly believed (see [\number\Papa]) that
$NP$-complete problems are drastically different from class-$P$ ones, at
least as far as the cardinality
of the continuum ({\sl i.e.} the number of points on the real line),
is drastically different from the cardinality of
natural numbers. Thus we argue that the class of
subminimal paths, {\sl i.e.} the paths close to the shortest
one, is drastically different from the class of the paths close to the
Hamilton path.

\section{4.~Two partial solutions}

\noindent
The following heuristic argument gives some insight on the ground state
of our model; but, in fact, it is easy to see that the argument should work
only in the non-frustrated case, {\sl i.e.} in the $MST$ problem.
Nevertheless we will see later that
the resulting relations fit the numerical data well enough
(see (\number\nftype) and next section). We try now
to estimate the length of the shortest path between $s$ and $t$.
Starting from $s$,
we select the outgoing arcs shorter than $\epsilon$ (to be determined),
and so on, until we reach $t$.
At each step the average number of selected outgoing arcs is
$$
N_\epsilon = N \; \int_0^\epsilon p_\alpha (x) \, dx
$$
where $ N = | V | $ is the number of nodes in $G$. After $n$ steps, the
number of reached nodes is about $ N_\epsilon^n $
(if $ N_\epsilon \gg 1 $). The probability of reaching $t$ in $n$ steps is
one if $ N_\epsilon^n = N $, {\sl i.e.}
$$
n = { \ln N \over \ln N_\epsilon }
$$
The length of the path thus constructed is approximately
$$
l = n \cdot { \int_0^\epsilon x p_\alpha (x) \, dx \over
       \int_0^\epsilon p_\alpha (x) \, dx }
$$
Taking the minimum as $ \epsilon $ varies, we get
$$
\eqalign{
n & = { \ln N \over \alpha + 1 } \cr
{ l \over n } & = { K \over N^{1/(\alpha+1)} } \qquad
\hbox{($ K = { \alpha + 1 \over \alpha + 2 } {\rm e} $)} \cr
}
\eqno (\number\nftype)
$$
We have performed a numerical analysis, measuring the average length of the
shortest path, and the average number of its arcs, for various values of $N$
and both uniform and linear probability distribution of the distances,
{\sl i.e.} for $\alpha=0$ and $\alpha=1$. The result of the analysis is
shown in next section: for the time being we stress that the numerical
data can be fitted with the following functional form, for $ N \gg 1 $
(see figures~1 and~2)
$$
\eqalign{
n & \approx c \cdot \ln N + C \cr
{ l \over n } & \approx { K \over N^s } \cr
}
\eqno (\number\nftype')
$$
This is in close agreement with (\number\nftype), and, fairly surprisingly,
also the numerical values of $c$ and $s$ agree with the ones in
(\number\nftype), {\sl i.e.} $ c = s \simeq 1/(\alpha+1) $. This fact will
be discussed in greater detail in the following.

\medskip

This approximate computation can be replaced with a more precise analysis
as follows. Consider the set $C_{s,t}$
of all the paths from $s$ to $t$, ordered by increasing length, and denote
with $ P^{(h)} ( n , \{ x_i \}_{i=1}^n ) $ the probability distribution
that the $h$-th path is composed of $n$ arcs of lengths
$ x_1, \dots, x_n $. We shall need the following notations: let $a_k$ be
the length of the arc $ k \in A $; given a path $ c \in C_{s,t} $,
let $l_c$ be its length, $n_c$ the number of its arcs, and
$ \{ k^i_c \}_{i=1}^{n_c} \subset A $ the sequence of arcs which it is made of
(so that $ l_c = \sum_{i=1}^{n_c} a_{k^i_c} $); finally we need the
characteristic function
$$
I_c^h \mkern 20mu = \sum_{c_1, \dots, c_{h-1}}
        \prod_{ c' \not= c_1, \dots, c_{h-1},\, c }
        \Theta(l_{c'} - l_c) \cdot
        \Theta(l_{c} - l_{c_{h-1}}) \cdots \Theta(l_{c_2} - l_{c_1})
\eqno (\number\Ihc)
$$
(which is 1 if $c$ is the $h$-th path, $0$ elsewhere). Then it is immediate
to see that:
$$
P^{(h)} ( n , \{ x_i \} ) = \int
   \bigl[
       \sum_c \,
       \delta_{n,n_c} \prod_{i=1}^n \delta(x_i - a_{k_c^i}) \, I^h_c
   \bigr] \prod_{k\in A} p_\alpha(a_k) \, d a_k
\eqno (\number\pexact)
$$
In order to make the computation of the integral in (\number\pexact)
feasible, we shall make the additional assumption
$$
\prod_{ k \in A } p_\alpha ( a_k ) =
\prod_{ c \in C_{s,t} } \prod_{i=1}^{n_c} p_\alpha ( a_{k^i_c} ) \; ,
\eqno (\number\approne)
$$
{\sl i.e.} two different paths do not have common arcs. Assumption
(\number\approne) should
be given a suitable sense ``in the average'' for $ N \gg 1 $ (since, as it
stands, it is evidently false).
However this is a minor problem, firstly
because we could not find better simplifying assumptions; secondly, because
we believe that to neglect the correlations between the paths (as in
(\number\approne)) corresponds to a kind of non-frustrated
approximation, which do not suitably fits our problem. We hope that
this point will become clearer in the following.
We omit here the cumbersome algebra, and give only the final results:
for $ N \gg 1 $ with $h$ fixed\footnote{${}^{(*)}$}
{$\gamma_E = 0.5772157 \dots $ is the Euler's gamma constant.}
$$
\eqalign{
\langle n \rangle^{(h)} & =
   { \ln N \over \alpha + 1 } + { 1 \over \alpha + 1 }
   \bigl[
      \ln ( \alpha + 1 ) + 1 - \gamma_E + \sum_{n=1}^{h-1} { 1\over n }
   \bigr] + O \bigl( { \ln N  \over N } \bigr) \cr
{ \langle l \rangle^{(h)} \over \langle n \rangle^{(h)} } & =
   { \alpha + 1 \over \chi \, N^{1/(\alpha+1)}} \;
   \bigl[
      1 - { 1 \over (\alpha+1) \langle n \rangle^{(h)} }
   \bigr]
   \qquad \chi = (\alpha + 1)!^{1/(\alpha+1)} \cr
}
\eqno (\number\lnresult)
$$
and
$$
{ \sigma^2_n \over \langle n \rangle^2 } = O \bigl( { 1 \over \ln N } \bigr)
\quad , \quad
{ \sigma^2_l \over \langle l \rangle^2 } = O \bigl( { 1 \over \ln^2 N } \bigr)
$$
where $ \langle n \rangle $ and $ \langle l \rangle $ denote the mean
value, respectively, of $n_c$ and $l_c$ in the distribution
(\number\pexact).

The relations (\number\lnresult) give to leading order the
non-frustrated behaviour of (\number\nftype). In other words, this more
refined computation with the simplifying assumption~(\number\approne) gives
again, as far as the leading behaviour is concerned, the result of the
previous heuristic argument, {\sl i.e.} a functional form of the
type~($\number\nftype'$) with $ c = s = 1/(\alpha+1) $, which, as we have
already pointed out, fits well enough the numerical data
(see next section and figures~1 and~2). An interesting feature of the
form~(\number\lnresult) is that the leading behaviour {\sl versus} $N$,
$ N \gg 1 $, of the number of arcs and of the length of the
$h$-th path, is not dependent on $h$: this is evident from
the fact that the dependence on $h$ in the right hand side of
the~(\number\lnresult) is only in the next-to-leading terms. This
feature is
very well confirmed by the numerical data (see again next section and
figures~1 and~2), so that, as we anticipated at the end of
previous section, we feel authorized to study all the subminimal paths
as a whole.
As far as the complete probability distribution of $l_c$ is concerned,
it results:
$$
\eqalign{
& P^{(h)} (l) = { H'_\alpha(l) \over (h-1)! } [ H_\alpha (l) ]^{h-1} \cdot
   \exp - H_\alpha (l) \cr
& H_\alpha (l) = { {\rm e}^{ {\cal N} l } \over N ( \alpha+1 ) } \qquad
   {\cal N} = (\alpha + 1)!^{1/(\alpha+1)} \, N^{1/(\alpha+1)} \cr
}
\eqno (\number\phl)
$$
The plots of the distributions $ P^{(h)} (l) $,
for $N=100$ and $ h = 1 $,~2 and~3, are shown in
figure~3, whereas in figure~5 a comparison with the experimental
results is exhibited. One can see that
the qualitative shapes of the distributions~(\number\phl) and the
experimental ones are rather
different, though their mean values are very close. For this reason we
believe that the assumption~(\number\approne), that we called
``non-frustrated'' and that was at the basis of
this computation, is not completely satisfactory to fit our problem.

\medskip

Another approach to the problem is
suggested by the cavity method of M\'ezard, Parisi and Virasoro
[\number\Cavity]. We can not pursue this approach beyond a kind of
non-frustrated approximation, to some extent equivalent to
the~(\number\approne) above, and very reminiscent of the
replica-symmetric approximation for the SK model.
We shall describe briefly
the procedure. Let $ G_N = ( V_N, A_N ) $ be a graph with $N$ nodes,
and fixed $s \in G_N$ let $\{ l_u \}_{u \in V_N \setminus \{ s \} }$ be the
lengths of the shortest paths in $G_N$ from $s$ to $u$. Let us denote with
$ {\cal P}_N ( \{ l_u \} ) $ their joint probability distribution. Let now
$ G_{N+1} = ( V_{N+1}, A_{N+1} ) $ be the graph with $N+1$ nodes obtained
by adding a node $u_0$ to $V_N$, and all the arcs between $u_0$ and the
nodes of $V_N$ to $A_N$; finally, let
$\{ {\cal L}_u \}_{u \in V_N}$ be the lengths of these new arcs, with
distribution $ P ( \{ {\cal L}_u \} ) = \prod p_\alpha ({\cal L}_u) $. The
shortest path from $s$ to $u_0$ in $G_{N+1}$ has length
$$
l_0 = \min [ {\cal L}_s , \min_{u \in V_N \setminus \{ s \} }
                    ( l_u + {\cal L}_u ) ]
\eqno (\number\lzero)
$$
As above, it is straightforward to write the exact form of the probability
distribution of $l_0$, but there is no point in using it
without suitable
simplifying assumptions. Namely, we shall assume the $ \{ l_u \} $ to be
uncorrelated, that is to say
$$
{\cal P}_N ( \{ l_u \} ) =
    \prod_{ u \in V_N \setminus \{ s \} } P_N ( l_u )
\eqno (\number\apprtwo)
$$
At first glance it is not clear which is the relation between this
approximation and the~(\number\approne) above: the latter
consists in neglecting the correlations among the paths between
the same pair of nodes, the first one
ignores the correlations between the shortest paths connecting a fixed
starting node with all the other ones of the graph.
However, as we shall see,
the resulting behaviour for the mean length of the shortest path is
exactly the same, though the two
complete probability distributions differ qualitatively.
Now, using the~(\number\apprtwo) one can translate the recursive
relation~(\number\lzero) in a recursive relation for the
probability distribution of the shortest path: writing
$$
F_N (l) = 1 - \int_0^l P_N (t) \, dt
$$
we obtain, in a more or less straightforward way,
$$
F_{N+1} (l) = \bigl[ 1 - \int_0^l p_\alpha \bigr] \cdot
   \exp - N \bigl(
               \int_0^l p_\alpha -
               \int_0^l p_\alpha ( l - x ) F_N ( x ) \, dx
            \bigr)
\eqno (\number\FNrel)
$$
If we now assume, apart from negligible terms,
$$
F^\infty_N (l) \simeq F_{N+1} (l) \simeq F_N (l) \qquad \hbox{for $N\gg1$,}
\eqno (\number\FNinfty)
$$
(\number\FNrel) becomes an integral (non-linear) equation in $F^\infty_N$.
It is possible to justify (\number\FNinfty) {\sl a posteriori}.

If $ \alpha = 0 $ the equation reduces to
$$
F^\infty_N (l) = (1-l) \cdot \exp - N ( l - \int_0^l F_N^\infty (x) \, dx )
$$
which is equivalent to
$$
\cases{
\displaystyle
{ d \over d l } F^\infty_N (l) =
   - { F^\infty_N \over 1-l } + F^\infty_N \cdot N [ F^\infty_N -1 ] & \cr
F^\infty_N (0) = 1 & \cr
}
\eqno (\number\FNzero)
$$
and can be linearized by putting $ T=1/F $. At the end we get
$$
F^\infty_N (l) = { 1 \over 1 +
  { \displaystyle \exp (Nl) - 1 \over \displaystyle N(1-l) }
  } \quad , \quad P^\infty_N (l) = - { d \over d l } F^\infty_N (l)
\eqno (\number\plzero)
$$
If $ \alpha \not = 0 $ the equation
(\number\FNrel), again under the hypothesis (\number\FNinfty), must be
solved numerically, thus giving
$ F^\infty_N (l) $ for each fixed value of $N$ and $\alpha$.
The resulting distributions $P^\infty_N(l)$ differs qualitatively from
the $P^{(h)}_N(l)$'s of equation~(\number\phl) above, even for $h=1$.
Nevertheless, as we shall see soon, their mean values coincide. A
comparison between these two distributions and the experimental one is shown
in figure~5, for $N=100$ and
both uniform and linear probability distribution of the distances,
{\sl i.e.} for $\alpha=0$ and $\alpha=1$.

As far as the mean value of $l$ is concerned, if $ \alpha = 0 $
the~(\number\plzero) can be easily integrated and gives
$$
\langle l \rangle = \int l P^\infty_N (l) \, d l \simeq { \ln N \over N }
\eqno (\number\lmean)
$$
in perfect
agreement with what is predicted by equation~(\number\lnresult).
If $ \alpha \not = 0 $ we must solve numerically the integral
equation~(\number\FNrel) for some selected values of $N$. Then we
compute $ \langle l \rangle $ and we
fit the resulting data with a functional form of the type
$$
\langle l \rangle = K' { \ln N \over N^s } + \hbox {subleading terms}
$$
We have performed
such an analysis for $\alpha = 0$ and $ \alpha=1$, computing
$ \langle l \rangle $ for the same values of $N$ of the experimental data,
thus giving to the finite-size effects the same relevance. The
details are given in next section. For $ \alpha = 0 $ we obtain
$$
K' = 1 \qquad s = 1
$$
with negligible corrections, in perfect agreement with~(\number\lmean),
thus showing that the selected values of $N$ are not too small. For
$ \alpha=1$, fairly
surprisingly, we have found again a very good
agreement with what is predicted by equation~(\number\lnresult), {\sl i.e.}
$$
K' = \sqrt{2}/2 \qquad s = 1/2
$$
Thus this
approach (suggested by the cavity method of [\number\Cavity] with the
assumption~(\number\apprtwo)), as to the leading behaviour
of the length of the shortest path,
is equivalent at all to the one which leads to
equation~(\number\lnresult), {\sl i.e.}
the direct one with the assumption~(\number\approne).

\medskip

As anticipated,
in next section we shall see that the non-frustrated behaviour of
equation~(\number\lnresult), shared by both our analytical approaches,
fits well enough the
experimental data for $ \langle l \rangle $ and $ \langle n \rangle $.
As we shall see, however,
the agreement is not beyond any doubts, so we believe that
it would be essential to improve the rough approximation (\number\apprtwo),
thus obtaining a more refined prediction than equation~(\number\lnresult).
An idea for a future work could be the following.

Let us assume that the Dijkstra's algorithm has given the $SPT$ for the
graph $G_N$
with $N$ nodes; then we dispose of the exact numerical values of the
lengths $l_u$ of the shortest paths from $s$ to any other node
$ u \in V_N $. If now we add the node $u_0$, obtaining $G_{N+1}$, the
length $l_0$ of the shortest path from $s$ to $u_0$ is given by
(\number\lzero). What
can be said now about the other lengths $ {\tilde l}_u $ of the shortest
paths from $s$ to the nodes of $V_N$? Of course, they will be affected by
the addition of $u_0$, in analogy to what happens in the SK model when
applying the cavity method. In fact the addition of one spin to an $N$-spin
system produces a rearrangement of the relative weights of the
configurations in a pure state, thence a minor rearrangement of the relative
weights of the pure states in a cluster and so on. The cavity method allows
to keep track of this cascade of rearrangements (see [\number\Cavity]
for details). The crucial observation was the discovery of
the cluster structure of the pure states in the SK model, the so called
{\sl ultrametric} organization (see [\number\Ultra]).
We do not yet know how to implement an
analogous procedure in our model, though it seems to be natural that an
improvement of the rough approximations
(\number\apprtwo) or (\number\approne) should be required.
In some sense the approximations we used here correspond to a
``replica-symmetric'' solution, which does not take into account the
rearrangement of the pure states. This would be exact if the system was
not frustrated ({\sl i.e.} with only one pure state, apart from accidental
degenerations).

The first condition, in order to start the program sketched
above, is to find out what plays the role of the SK pure states in our
model. Let us therefore consider the following argument. After adding the
node $u_0$ to the graph $G_N$ obtaining $G_{N+1}$, one can apply the
Dijkstra's
algorithm to find the $SPT$ in $G_{N+1}$. Take now a node $u$ in $V_N$. If
the shortest path in $G_{N+1}$ from $s$ to $u$ does not pass trough $u_0$,
then it will coincide with the old shortest path in $G_N$.
If, on the contrary, it
passes through $u_0$, it is interesting to see how much ``close'' (in a
suitable sense) it is to the old shortest path in $G_N$. In general, there
is no reason to expect these two paths to be close, because of the
frustration of the model, but one can reasonably suppose that the new path
runs close to an old subminimal path in $G_N$ from $s$ to $u$.
Therefore in our numerical analysis we have pointed out our interest on the
suboptimal paths a little longer than the minimal one: can we interpret the
suboptimal paths as the pure states (in a suitable sense) of our model?
Let us stress that to leading order for $N\gg1$ the lengths
$ \langle l \rangle^{(h)} $ of the suboptimal paths coincide, as it
results from
(\number\lnresult) and is confirmed by numerical analysis
(see next section). Otherwise
the identification between subminimal paths and pure states will fail:
in the SK model the free energies are equal in the thermodynamical limit.

\section{5.~Numerical results}

\noindent
As pointed out above, we have performed some
numerical computations to test the validity of the
scaling laws (\number\nftype), and to measure the probability distribution
for the lengths of the subminimal paths. The algorithm used is a clever
generalization of the Dijkstra's algorithm [\number\PDG].
It determines, for a given realization of the
distance matrix $d(u,v)$, the first $k$ paths of $C_{s,t}$ ordered by
increasing
length. We have considered only $k=3$, because the CPU-time
spent in the computation grows rather fast with $k$.
We made several runs for $\alpha = 0 $ and $ \alpha = 1 $, {\sl i.e.}
with the
probability distribution in (\number\defp) respectively uniform and linear,
and for different values of $N$, the size of the
graph, ranging from $N=10$ to $N=500$. In each run we measure the
length and the number of arcs of the paths for a large number of
independent
realizations of the distance matrix. More precisely we made
$200.000$ iterations for each value of $\alpha $
for $ N = 10, 20, 50, 100 $; $100.000$ iterations for $ N = 200 $;
$ 40.000 $ iterations for $ N = 500 $.
The runs required about 150 hours of CPU on a DEC VAX 6000-520,
running VAX Pascal. The runs to measure the probability distribution, being
purely qualitative and not requiring a high precision, were made only for
$N=100$ and low statistics:
$30.000$ iterations for each value of $\alpha$.

\medskip

In figure~1 and~2 we plot respectively
$ \langle n \rangle^{(h)} $ and
$ \langle l \rangle^{(h)} / \langle n \rangle^{(h)} $ {\sl versus} the size
of the graph $N$, for $h=1,2,3$ and both probability distributions for the
distances ($\alpha=0$ and $\alpha=1$). The scaling laws that qualitatively
appear in these plots are the ones anticipated in
equation~($\number\nftype'$), {\sl i.e.} for $N \gg 1 $
$$
\eqalign{
\langle n \rangle^{(h)} & \approx c \cdot \ln N + C^{(h)} \cr
{ \langle l \rangle^{(h)} \over \langle n \rangle^{(h)} }
                        & \approx { K \over N^s }
                        \quad \hbox{independent on $h$} \cr
}
$$
We want to stress once more that to
leading order in $N$ these scaling laws do not depend on $h$.

To be sure that the leading behaviour for the ``physical'' quantities
$\langle n \rangle$ and $\langle l \rangle$ is the non-frustrated one of
equation~(\number\nftype) and~(\number\lnresult), we
should be more quantitative, and show that
$$
c = s = {1 \over \alpha + 1 }
\eqno (\number\csrel)
$$
We cannot simply perform a na\"\i ve least-squares fit, as our
data could not have reached their asymptotic behaviour, because of the low
involved values of $N$. Therefore our data have been fitted with curves of the
form
$$
\eqalign{
\langle n \rangle & = c \cdot \ln N + C + B { \ln N \over N } \cr
{ \langle l \rangle \over \langle n \rangle } & = { K \over N^s }
                  \bigl( 1 + { B \over \ln N } \bigr) \cr
}
\eqno (\number\fit)
$$
for various {\sl fixed} values of $B$, applying the {\sl flatness criterion}
advocated in [\number\BS, sections~4.2 and~5.3]. The $B$-dependent term
should take into account the corrections to the leading behaviour,
and its form was suggested by equation~(\number\lnresult).
Although we believe that this technique has been now generally accepted,
in the appendix we shall spend some words of explanation.
The result of the analysis for
$\langle n \rangle$ and $\langle l \rangle / \langle n \rangle$
is given in table~1 and~2. From these data we see that
relation~(\number\csrel) is verified well enough. Sufficiently well verified
is also the fact that $K$ is independent on $h$, at least for
$\alpha=1$. As to the value of $C^{(h)}$ the agreement with
equation~(\number\lnresult) is only qualitative.

Nevertheless the fitted values for $c$ and $s$ in table~1 and~2 are
different
from what is predicted by equation~(\number\csrel) by more than their
error, though this should be a 95\% confidence limit. Thus we conclude
that, as we anticipated at the end of the previous section,
the use of the non-frustrated leading behaviour~(\number\lnresult) to
fit our experimental data is not completely satisfactory.
We strongly believe that the finite size errors are not
underestimated, because the same procedure of fitting
performed on the theoretical data leads to confirm the validity
of~(\number\csrel) exactly, {\sl i.e.} at 95\% confidence level.
More precisely we have proceeded as follows. We have generated a set of
values for $ \langle l \rangle $ integrating the theoretical
distributions of equation~(\number\phl) and~(\number\FNrel), for the same
values of $N$ of the real data,
and with fictitious statistical errors equal to the real ones. Then we have
performed in parallel the same fit, with the same procedure of the previous
analysis, on
the experimental data and the theoretical ones,
which so work as a {\sl placebo}. The result is
given in table~3: the data have been fitted with the functional form
$$
\langle l \rangle = K' { \ln N \over N^s }
                  \bigl( 1 + { B \over \ln N } \bigr)
\eqno (\number\fit')
$$
where again
the corrections are suggested by equation~(\number\lnresult).
We see that the theoretical data give the expected values
well within their error, with some troubles for $h=1$, not only for
the exponent
$ s $ ({\sl i.e.} $ 1/(\alpha+1) $), but also for the constant
$ K' $ ({\sl i.e.} $1$ for $ \alpha = 0 $ and
$ \sqrt{2}/2 $ for $ \alpha = 1 $, see~(\number\lnresult)).
As to the experimental data, though
the flatness region in $B$ is larger, thus producing a systematic error
larger, the fitted value for
$ s $ differs from $ 1/ (\alpha+1) $ by more than its error, and
the discrepancy becomes stronger when $ h $ increases.

\medskip

In figure~3, 4~and~5 we show the various
probability distributions for the length of the subminimal paths for
$ N = 100 $ and both $\alpha=0$ and $\alpha=1$.
In figure~3 we plotted the distributions
$ P^{(h)} (l) $ of equation~(\number\phl) for $ h = 1$, 2~and~3.
Figure~4 presents the experimental distributions for $ h = 1$, 2~and~3,
obtained by smoothing the histogram of the experimental data. More
precisely we filled a histogram with the experimental data for
$ l $, for different realizations of the distance matrix (200 channels
for $30.000$ data, no one of the data falls out of the considered range),
then we averaged the counting of each channel with
the $H$ preceeding and the $H$ following it
($ H = 5 $ for $ \alpha = 0 $ and $ H = 2 $ for $ \alpha = 1 $), thus
obtaining a smoother plot.
Lastly, as far as the length of the shortest path is concerned,
in figure~5 we compare
the distribution $ P^{(h)} (l) $ with $ h = 1 $ of figure~3,
the experimental distribution of figure~4,
and the distribution of the ``cavity method'', {\sl i.e.} the one
obtained by solving equation~(\number\FNrel), exactly for $ \alpha = 0 $
(see~(\number\plzero)), and numerically for $ \alpha = 1 $.

As a comment to figure~5 we remark again (this fact has been yet noted in
the previous section) that the two theoretical distributions and the
experimental one differ qualitatively from each others,
at $N$ fixed, though the first two distributions
share the same leading behaviour {\sl versus} $N$ as to their mean
value, whereas the third one gives a
significatively different behaviour.

\section{Appendix}

\noindent
In this appendix we sketch briefly the contents of the analysis of the
corrections to the scaling {\sl \`a la} Berretti-Sokal~[\number\BS].
In our case we
measure $ \langle l \rangle $ and $ \langle n \rangle $ for some values of
$N$, the size of the graph, and we would extrapolate from the data
the scaling behaviour {\sl versus} $N$. Because of the involved relatively
small values of $N$, we can not be sure that the data have reached their
asymptotic behaviour, so we must take into account possible corrections to
the scaling. Firstly we make an Ansatz for
the corrections to the leading behaviour, as in~(\number\fit).
The coefficient $B$ of the corrections in~(\number\fit) is {\sl not}
subject to a fit, although
this would be easily made by well known non-linear methods of fitting,
because this would not solve our problem, {\sl i.e.} to understand
to what extent the fitted parameters are reliable as true asymptotic
values, and not merely as effective
values, changing as the range in $N$ increases. In fact we do not
know exactly which the subleading corrections are, and also, if we knew them,
we could not include in a fit other but the first relevant ones, as our
data are only for a few values of $N$ (6, in our case).

Thus we fix the value of $B$, we fix also a value $N_{cut}$ for $N$, and we
fit the data with the curves in~(\number\fit) only for $ N > N_{cut} $.
Varying $N_{cut}$ at $B$ fixed, if we observe a systematic dependence on
$N_{cut}$ of the fitted parameters ({\sl e.g.} a systematic decrease
of the fitted exponent $s$), we say that the chosen value of $B$ can not
take into account the corrections to the scaling. Now, by varying $B$,
we select the range in $B$ for which the fitted parameters are flat
({\sl i.e.} do not vary in a systematic way), with respect to
$N_{cut}$. We say, following~[\number\BS], that for $B$ in this range
equation~(\number\fit) takes into account the corrections to the scaling as
an effective correction, even if the exact form for the corrections is
different from that we imposed, and even if we rule out some other (more
irrelevant) terms. In this range of $B$, which we call the
``flatness region'', for each fitted parameter we select the maximum and
the minimum value. The best fit for the parameter will be simply the
arithmetic mean of these values, whereas their difference will be
considered as the systematic error due to unconsidered corrections to
the scaling, or to imperfect knowledge of the form of the corrections
(95\% subjective confidence limit as defined in
[\number\BS, footnote 17]). In addition we quote the usual statistical
error for the fit, at 95\% confidence level (2$\sigma$).


\section{Acknowledgements}

\noindent
I am greatly indebted to Giorgio Parisi
for having originally proposed this work, and
for many stimulating discussions
and helpful suggestions throughout all its development.
I am also grateful to Sergio Caracciolo and Andrea Pelissetto for their
kind collaboration in some aspects of this work.
All numerical computations were performed at Scuola Normale Superiore in
Pisa.


\section{References}

\parskip=5pt

\item{[\number\SK]} \author {D.Sherrington and S.Kirkpatrick} \hfill\break
      \revtit {Solvable model of a spin glass}
      {Phys.Rev.Lett.} {35} {1792} {1975}

\item{[\number\Parisi]} \author {G.Parisi}
      \revtit {A sequence of approximated solutions to the SK model for
               spin glasses}
      {J.Phys.A} {13} {L115} {1980} \hfill\break
      \revtit {The order parameter for spin glasses: a function on the
               interval $[0,1]$}
      {J.Phys.A} {13} {1101} {1980} \hfill\break
      \revtit {Magnetic properties of spin glasses in a new mean field
               theory}
      {J.Phys.A} {13} {1887} {1980}

\item{[\number\Overlap]} \author {G.Parisi}
      \revtit {Order parameter for spin glasses}
      {Phys.Rev.Lett.} {50} {1946} {1983}

\item{[\number\Ultra]} \author {M.M\'ezard, G.Parisi, N.Sourlas, G.Toulouse
                                and M.A.Virasoro}
      \revtit {Replica symmetry breaking and the nature of the spin glass
               phase}
      {J.Physique} {45} {843} {1984}

\item{[\number\RFE]} \author {M.M\'ezard, G.Parisi and M.A.Virasoro}
      \revtit {Random free energies in spin glasses}
      {J.Physique Lettres} {46} {L217} {1985}

\item{[\number\Cavity]} \author {M.M\'ezard, G.Parisi and M.A.Virasoro}
      \revtit {SK model: the replica solution without replicas}
      {Europhys.Lett.} {1} {77} {1986}

\item{[\number\Ruelle]} \author {D.Ruelle}
      \revtit {A mathematical reformulation of Derrida's REM and GREM}
      {Commun.Math.Phys.} {108} {225} {1987}

\item{[\number\CPPS]} \author {S.Caracciolo, G.Parisi, S.Patarnello and
                               N.Sourlas}
      \revtit {Low temperature behaviour of 3-D spin glasses in a magnetic
               field}
      {J.Physique} {51} {1877} {1990}

\item{[\number\FH]} \author {D.S.Fisher and D.A.Huse} \hfill\break
      \revtit {Ordered phase of short-range Ising spin glasses}
      {Phys.Rev.Lett.} {56} {1601} {1986}

\item{[\number\REM]} \author {B.Derrida}
      \revtit {Random-energy model: an exactly solvable model of disordered
               systems}
      {Phys.Rev.B} {24} {2613} {1981}

\item{[\number\GREM]} \author {B.Derrida}
      \revtit {A generalization of the random energy model which includes
               correlations between energies}
      {J.Physique Lettres} {46} {L401} {1985}

\item{[\number\Papa]} \author {C.H.Papadimitriou and K.Steiglitz}
      \revbook {Combinatorial optimization: algorithms and complexity}
      {Prentice-Hall, Inc.} {1982}

\item{[\number\Barahona]} \author {F.Barahona}
      \revtit {On the computational complexity of Ising spin glass models}
      {J.Phys.A} {15} {3241} {1982}

\item{[\number\dyrpol]} \author {M.M\'ezard and G.Parisi}
      \revtit {Interfaces in a random medium and replica symmetry breaking}
      {J.Phys.A} {23} {L1229} {1990}

\item{[\number\PDG]} \author {P.Digianantonio} private communication.

\item{[\number\BS]} \author {A.Berretti and A.D.Sokal}
      \revtit {New Montecarlo method for the self-avoiding walk}
      {J.Stat.Phys.} {40} {483} {1985}

\vfill \eject


\nopagenumbers

\section{Captions for the figures and tables}

\noindent
{\bf Figure 1.} $ \langle n \rangle^{(h)} $ {\sl versus} $N$,
with uniform ($a$) and linear ($b$)
probability distribution for the distances.

\noindent
{\bf Figure 2.} The same as in figure~1 for
$ \langle l \rangle^{(h)} / \langle n \rangle^{(h)} $.

\noindent
{\bf Figure 3.} Distributions $ P^{(h)} (l) $ for the length of the $h$-th
path as in equation~(\number\phl), for $N=100$ and
uniform ($a$) and linear ($b$)
probability distribution for the distances.

\noindent
{\bf Figure 4.} The same as in figure~3 for the
measured distributions $ P_{exp}^{(h)} (l) $,
obtained by smoothing the histogram of the experimental data.

\noindent
{\bf Figure 5.} Three probability distributions for the length of the
shortest path, with $N=100$ and
uniform ($a$) and linear ($b$)
probability distribution for the distances.
$ P^{(1)} (l) $ is the distribution quoted in (\number\phl), as in
figure~3, for $h=1$;
{\tt exp} denotes the experimental distribution, as in figure~4, for $h=1$;
{\tt cav} denotes the distribution
of the ``cavity method'' (see equation~(\number\FNrel)).

\noindent
{\bf Table 1.} Best estimates for $c$ and $C$ in the fit
$ \langle n \rangle^{(h)} = c \cdot \ln N + C $, with corrections to the
leading behaviour inserted (see~(\number\fit)).
The first error is the systematic error due to unconsidered corrections
to the scaling (95\% subjective confidence limit as defined in
[\number\BS, footnote 17]) and the second error is the usual
statistical error (95\% confidence interval). It is also quoted the range
in $B$ which obeys to the flatness criterion.

\noindent
{\bf Table 2.} Best estimates as in table~1 for $s$ and $K$ in the fit
$ \langle l \rangle^{(h)} / \langle n \rangle^{(h)} = K / N^s $,
with corrections to the leading behaviour inserted (see~(\number\fit)).

\noindent
{\bf Table 3.} Best estimates as in table~1 for $s$ and $K'$ in the fit
$ \langle l \rangle^{(h)} = K' \cdot \ln N / N^s $,
with corrections to the leading behaviour inserted (see~($\number\fit'$)).
The fit has been performed for both the experimental
data (denoted by {\tt exp}) and the theoretical ones. These last were
produced by integrating the two theoretical probability distributions for $l$,
the one of equation~(\number\phl) and the one of the cavity
equation~(\number\FNrel).

\vfill \eject


\line{
  \hfil {\bf Figure~1.a}}

\setbox\BodyBox = \vbox{
   \epsfysize = 0.8\hsize
   \setbox\RotateBox=\hbox{\epsfbox[25 10 550 720] {fig1a.ps}}
   \rotr\RotateBox}

\centerline{\box\BodyBox}

\vfill

\line{
  \hfil {\bf Figure~1.b}}

\setbox\BodyBox = \vbox{
   \epsfysize = 0.8\hsize
   \setbox\RotateBox=\hbox{\epsfbox[25 10 550 720] {fig1b.ps}}
   \rotr\RotateBox}

\centerline{\box\BodyBox}

\vfill \eject

\line{
  \hfil {\bf Figure~2.a}}

\setbox\BodyBox = \vbox{
   \epsfysize = 0.8\hsize
   \setbox\RotateBox=\hbox{\epsfbox[25 10 550 720] {fig2a.ps}}
   \rotr\RotateBox}

\centerline{\box\BodyBox}

\vfill

\line{
  \hfil {\bf Figure~2.b}}

\setbox\BodyBox = \vbox{
   \epsfysize = 0.8\hsize
   \setbox\RotateBox=\hbox{\epsfbox[25 10 550 720] {fig2b.ps}}
   \rotr\RotateBox}

\centerline{\box\BodyBox}

\vfill \eject

\line{
  \hfil {\bf Figure~3.a}}

\setbox\BodyBox = \vbox{
   \epsfysize = 0.8\hsize
   \setbox\RotateBox=\hbox{\epsfbox[25 10 550 720] {fig3a.ps}}
   \rotr\RotateBox}

\centerline{\box\BodyBox}

\vfill

\line{
  \hfil {\bf Figure~3.b}}

\setbox\BodyBox = \vbox{
   \epsfysize = 0.8\hsize
   \setbox\RotateBox=\hbox{\epsfbox[25 10 550 720] {fig3b.ps}}
   \rotr\RotateBox}

\centerline{\box\BodyBox}

\vfill \eject

\line{
  \hfil {\bf Figure~4.a}}

\setbox\BodyBox = \vbox{
   \epsfysize = 0.8\hsize
   \setbox\RotateBox=\hbox{\epsfbox[25 10 550 720] {fig4a.ps}}
   \rotr\RotateBox}

\centerline{\box\BodyBox}

\vfill

\line{
  \hfil {\bf Figure~4.b}}

\setbox\BodyBox = \vbox{
   \epsfysize = 0.8\hsize
   \setbox\RotateBox=\hbox{\epsfbox[25 10 550 720] {fig4b.ps}}
   \rotr\RotateBox}

\centerline{\box\BodyBox}

\vfill \eject

\line{
  \hfil {\bf Figure~5.a}}

\setbox\BodyBox = \vbox{
   \epsfysize = 0.8\hsize
   \setbox\RotateBox=\hbox{\epsfbox[25 10 550 720] {fig5a.ps}}
   \rotr\RotateBox}

\centerline{\box\BodyBox}

\vfill

\line{
  \hfil {\bf Figure~5.b}}

\setbox\BodyBox = \vbox{
   \epsfysize = 0.8\hsize
   \setbox\RotateBox=\hbox{\epsfbox[25 10 550 720] {fig5b.ps}}
   \rotr\RotateBox}

\centerline{\box\BodyBox}

\vfill \eject


\newbox\mystrutbox
\setbox\mystrutbox=\hbox{\vrule height15pt width0pt depth5pt}
\def\mystrut{\relax\ifmmode\copy\mystrutbox\else\unhcopy\mystrutbox\fi}

\line{}
\vfill

\line{
  \hfil {\bf Table~1.}}
$$ \vbox { \offinterlineskip \tabskip=0pt
\halign {%
\mystrut \vrule # &
$ # $ \hss & \quad \vrule # &
\quad \hss $ # $ \hss &
\quad \hss $ # $ \hss &
\quad \hss $ # $ \hss \vrule \cr
\noalign {\hrule}
&            & &  c                          & C                  & B \cr
\noalign {\hrule}
& \; \alpha = 0 & &                          &                    &   \cr
& \mkern20mu h = 1 & & 1.0070 \pm 0.0079 \pm 0.0023 &
                       -0.475 \pm 0.041  \pm 0.008  &  0.87 \div  0.99 \cr
& \mkern20mu h = 2 & & 1.0363 \pm 0.0098 \pm 0.0024 &
                        0.316 \pm 0.051  \pm 0.009  & -0.20 \div -0.05 \cr
& \mkern20mu h = 3 & & 1.0627 \pm 0.0179 \pm 0.0125 &
                        0.650 \pm 0.095  \pm 0.060  & -0.60 \div -0.40 \cr
\noalign {\hrule}
& \; \alpha = 1 & &                          &                    &   \cr
& \mkern20mu h = 1 & & 0.5051 \pm 0.0108 \pm 0.0025 &
                        0.283 \pm 0.061  \pm 0.010  & -0.55 \div -0.30 \cr
& \mkern20mu h = 2 & & 0.5051 \pm 0.0087 \pm 0.0074 &
                        0.781 \pm 0.051  \pm 0.036  & -0.50 \div -0.30 \cr
& \mkern20mu h = 3 & & 0.5138 \pm 0.0129 \pm 0.0026 &
                        0.971 \pm 0.073  \pm 0.011  & -0.30 \div  0.00 \cr
\noalign {\hrule}
}}$$

\vfill

\line{
 \hfil {\bf Table~2.}}
$$ \vbox { \offinterlineskip \tabskip=0pt
\halign {%
\mystrut \vrule # &
$ # $ \hss & \quad \vrule #
& \quad \hss $ # $ \hss
& \quad \hss $ # $ \hss
& \quad \hss $ # $ \hss \vrule \cr
\noalign {\hrule}
&            & &  s                          & K                  & B \cr
\noalign {\hrule}
& \; \alpha = 0 & &                          &                    &   \cr
& \mkern20mu h = 1 & & 1.0191 \pm 0.0118 \pm 0.0021 &
                        1.246 \pm 0.131  \pm 0.012  & 0.25 \div 0.50 \cr
& \mkern20mu h = 2 & & 0.9914 \pm 0.0106 \pm 0.0016 &
                        0.844 \pm 0.094  \pm 0.006  & 1.70 \div 2.10 \cr
& \mkern20mu h = 3 & & 0.9780 \pm 0.0070 \pm 0.0032 &
                        0.675 \pm 0.049  \pm 0.010  & 2.90 \div 3.20 \cr
\noalign {\hrule}
& \; \alpha = 1 & &                          &                    &   \cr
& \mkern20mu h = 1 & & 0.4981 \pm 0.0102 \pm 0.0018 &
                        1.361 \pm 0.119  \pm 0.011  & -0.08 \div  0.10 \cr
& \mkern20mu h = 2 & & 0.4920 \pm 0.0076 \pm 0.0013 &
                        1.265 \pm 0.084  \pm 0.008  &  0.20 \div  0.35 \cr
& \mkern20mu h = 3 & & 0.4948 \pm 0.0076 \pm 0.0012 &
                        1.290 \pm 0.086  \pm 0.007  &  0.20 \div  0.35 \cr
\noalign {\hrule}
}}$$

\vfill\eject

\line{
 \hfil {\bf Table~3.}}
$$ \vbox { \offinterlineskip \tabskip=0pt
\halign {%
\mystrut \vrule # &
$ # \mkern10 mu $ \hss & \vrule #
& \quad \hss $ # $ \hss
& \quad \hss $ # $ \hss
& \quad \hss $ # $ \hss \vrule \cr
\noalign {\hrule}
&            & &  s                          & K'                 & B \cr
\noalign {\hrule}
& \; \alpha = 0 & &                          &                    &   \cr
& \mkern10mu {\tt exp} \; h = 1 & & 1.0077 \pm 0.0077 \pm 0.0013 &
                        1.098 \pm 0.074  \pm 0.007  & 0.20 \div 0.35 \cr
& \mkern10mu {\tt exp} \; h = 2 & & 0.9874 \pm 0.0081 \pm 0.0008 &
                        0.814 \pm 0.075  \pm 0.003  & 2.50 \div 2.90 \cr
& \mkern10mu {\tt exp} \; h = 3 & & 0.9667 \pm 0.0044 \pm 0.0005 &
                        0.571 \pm 0.034  \pm 0.001  & 5.50 \div 5.90 \cr
& \mkern10mu eq.~(\number\FNrel) & & 1 &  1 &       0        \cr
& \mkern10mu eq.~(\number\phl) \; h = 1 & & 0.9935 \pm 0.0038 \pm 0.0007 &
                        0.932 \pm 0.030  \pm 0.003  & -0.36 \div -0.43 \cr
& \mkern10mu eq.~(\number\phl) \; h = 2 & & 0.9998 \pm 0.0011 \pm 0.0003 &
                        0.997 \pm 0.009  \pm 0.001  & 0.415 \div 0.445 \cr
& \mkern10mu eq.~(\number\phl) \; h = 3 & & 0.9999 \pm 0.0013 \pm 0.0002 &
                        0.999 \pm 0.012  \pm 0.001  & 0.910 \div 0.935 \cr
\noalign {\hrule}
& \; \alpha = 1 & &                          &                    &   \cr
& \mkern10mu {\tt exp} \; h = 1 & & 0.5063 \pm 0.0041 \pm 0.0007 &
                        0.763 \pm 0.027  \pm 0.002  & 0.16 \div 0.24 \cr
& \mkern10mu {\tt exp} \; h = 2 & & 0.4990 \pm 0.0035 \pm 0.0004 &
                        0.692 \pm 0.025  \pm 0.001  & 1.51 \div 1.63 \cr
& \mkern10mu {\tt exp} \; h = 3 & & 0.4931 \pm 0.0033 \pm 0.0004 &
                        0.638 \pm 0.023  \pm 0.001  & 2.40 \div 2.55 \cr
& \mkern10mu eq.~(\number\FNrel) & & 0.4994 \pm 0.0025 \pm 0.0006 &
                        0.702 \pm 0.014  \pm 0.002  & 0.29 \div 0.33 \cr
& \mkern10mu eq.~(\number\phl) \; h = 1 & & 0.4951 \pm 0.0025 \pm 0.0006 &
                        0.670 \pm 0.013  \pm 0.002  & 0.25 \div 0.29 \cr
& \mkern10mu eq.~(\number\phl) \; h = 2 & & 0.4999 \pm 0.0016 \pm 0.0002 &
                        0.706 \pm 0.010  \pm 0.001  & 1.10 \div 1.14 \cr
& \mkern10mu eq.~(\number\phl) \; h = 3 & & 0.5000 \pm 0.0010 \pm 0.0001 &
                        0.707 \pm 0.007  \pm 0.001  & 1.60 \div 1.63 \cr
\noalign {\hrule}
}}$$

\bye